\documentclass[11pt, a4paper]{article}
\usepackage{jheppub}
\usepackage{graphicx}
\usepackage{epsfig,amsmath,amsthm}
\newcommand{\be}{\begin{equation}}
\newcommand{\ee}{\end{equation}}
\newcommand{\bea}{\begin{eqnarray}}
\newcommand{\eea}{\end{eqnarray}}
\def\bse{\begin{subequations}}
\def\ese{\end{subequations}}
 
\newcommand{\IN}{\mathbb{N}}
\def\IZ{\relax\ifmmode\hbox{Z\kern-.4em Z}\else{Z\kern-.4em Z}\fi}

\newcommand{\non}{\nonumber \\}

\def\half{\frac{1}{2}} 

\def\del{{\partial}}

\def\hd{{\hat d}} \def\hl{{\hat l}}
\def\hlam {{\hat \lambda}} \def\hy{{\hat Y}}

 \def\bphi{{\bar \phi}}
 
\def\cl{{\cal L}} \def\co{{\cal O}}

\def\al{\alpha} 
  \def\eps{\epsilon}

\def\lam{\lambda}

\def\presub{\vspace{.5cm} \noindent}

\def\bi{\begin{itemize}} \def\ei{\end{itemize}}

\def\Schw{Schwarzschild }
\def\({\left(} \def\){\right)}
\def\[{\left[} \def\]{\right]}

\title{ \center{Black hole stereotyping:\\
 Induced gravito-static polarization}}

\author{Barak Kol\\
{\it Racah Institute of Physics, Hebrew University,\\
 Jerusalem 91904, Israel} \\
{\tt\href{mailto:barak_kol@phys.huji.ac.il}{barak\_kol@phys.huji.ac.il}}\\
\vspace{.3cm}
 Michael Smolkin\\
{\it Perimeter Institute for Theoretical Physics,\\
 Waterloo, Ontario N2L 2Y5, Canada} \\
 {\tt\href{mailto:msmolkin@perimeterinstitute.ca}{msmolkin@perimeterinstitute.ca}}
 }

\abstract{We discuss the black hole effective action and define
its static subsector. We determine the induced gravito-static
polarization constants (electric Love numbers) of static black
holes (Schwarzschild) in an arbitrary dimension, namely the
induced mass multipole as a result of an external gravitational
field. We demonstrate that in 4d these constants vanish thereby
settling a disagreement in the literature. Yet in higher
dimensions these constants are non-vanishing, thereby disproving
(at least in $d>4$) speculations that black holes have no
effective couplings beyond the point particle action. In
particular, when $l/(d-3)$ is half integral these constants
demonstrate a (classical) renormalization flow consistent with the
divergences of the effective field theory. In some other cases the
constants are negative indicating a novel non-spherical
instability.
 The theory of hypergeometric functions plays a central role.}

\begin{document}
\maketitle

\section{Introduction}

\begin{figure}[t]
\begin{center}
\includegraphics[width=11cm]{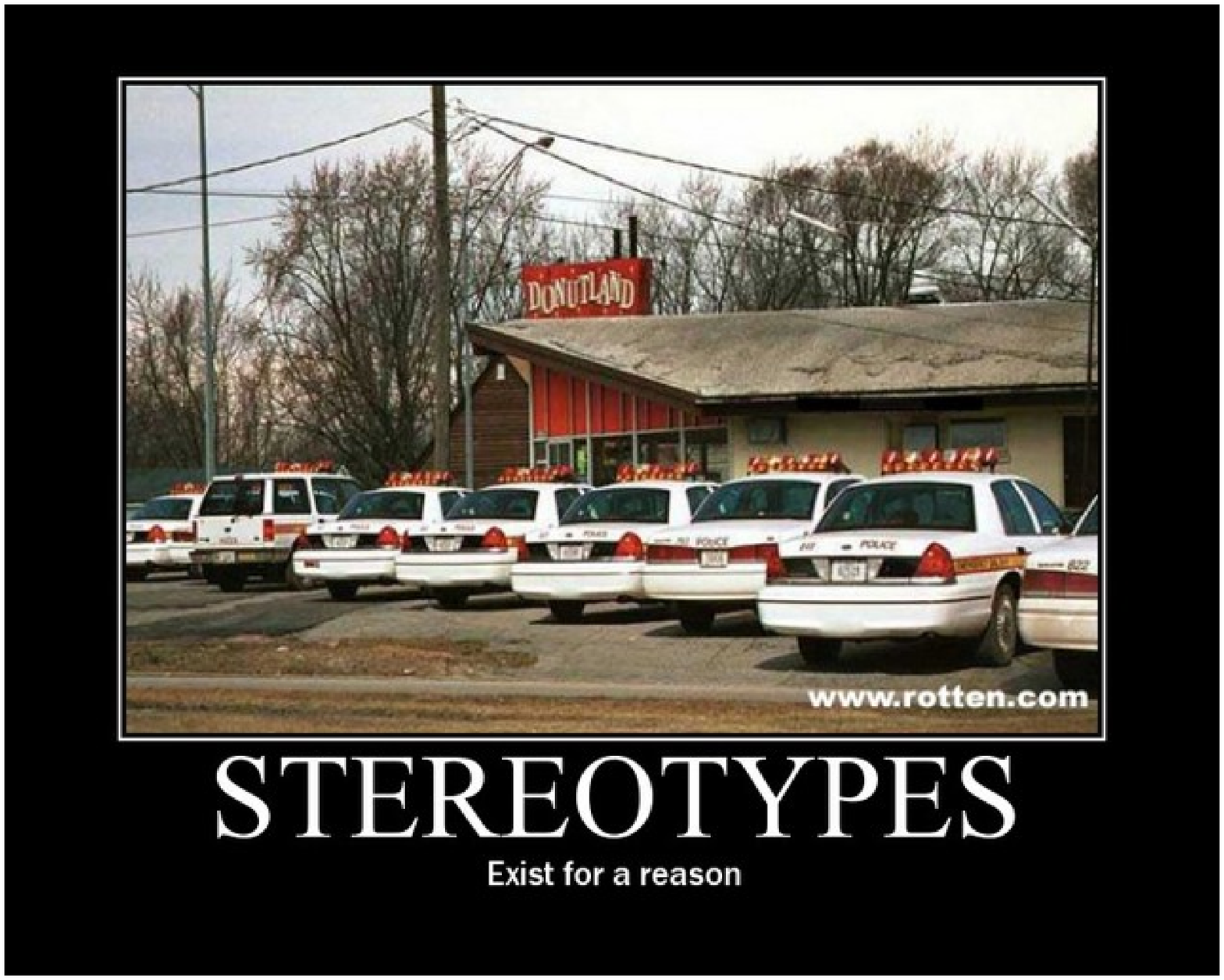}
\caption{}
\label{fig:stereo}
\end{center}
\end{figure}

The word ``stereotype'' carries a negative connotation for judging
an individual based on a group to which he or she belongs rather
than basing on actual individual traits. \footnote{ Stereotype
(dictionary definition \cite{oed}) -- A preconceived and
oversimplified idea of the characteristics which typify a person,
situation, etc.} Yet, in this work we shall describe a useful
treatment of compact objects including black holes which can be
considered as stereotyping
(see figure \ref{fig:stereo}).

Most of this paper is devoted to a determination of the
coefficients of induced gravito-static polarization, known as the
electric Love numbers \cite{Love},
\footnote{Named after the British mathematician A. E. H. Love who defined them first in 1911 in the context of Newtonian gravity.}
 of a black hole in 4d and in higher
dimensions. Consider a non-trivial gravito-static space-time and
consider placing a small non-spinning black hole at an equilibrium
point. As a result of the tidal gravitational force the black hole
will deform, mass multipole moments will be measurable from far
away, and could be interpreted as changes in the internal mass
distribution of the black hole. To leading order the induced
multipole moment is proportional to the external field multipole
at the equilibrium point and we shall refer to the constant of
proportionality as the coefficient of induced gravito-static
polarization, or Love number.

\begin{figure}[t]
\begin{center}
\includegraphics[width=10cm]{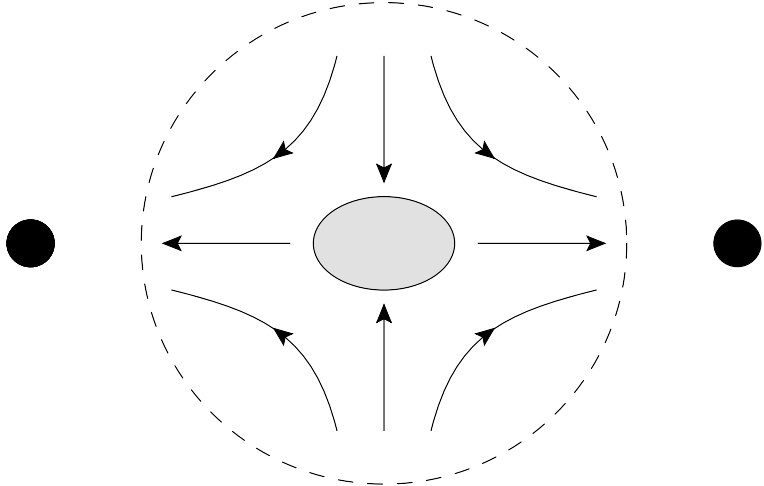}
\caption{Experimental set-up for measuring the induced
gravito-static polarization constants. The black hole (gray) is
positioned at an equilibrium point between two fixed stars
(black). Its mass multipole moments are measured on a far away
envelope (dashed sphere).} \label{fig:setup}
\end{center}
\end{figure}

As a concrete example, consider two very massive and approximately
static bodies (see figure \ref{fig:setup}) . On the line joining
them there is a point of gravito-static equilibrium. We place a
small BH at that point and wish to measure the induced mass
multipole moments from far away. Another concrete example, one
which is exactly static, is a caged black hole, namely a small
black hole within a larger extra dimension (see for example  \cite{Chu:2006ce},\cite{CLEFT-caged},\cite{Gilmore:2009ea} and reference therein). In this
case the black hole reacts to a gravitational field created by
itself (through its ``mirror images'').

These constants are basic linear response coefficients for black
hole physics, much like compressibility in elasticity. Apart for
the intrinsic interest, knowing these constants would have several
applications. First, they are needed for solving the binary
motion. While for 4d black holes such effects appear first (in the
two body effective action) only at order 5PN, they are much more
relevant for other types of compact objects and moreover in very high dimension the leading effect for BHs appears already at order
$1^+$PN, namely only a little over 1PN. Another possible application is the study of extended black objects in higher
dimensions and their instabilities, as we shall find out.

Even though black hole Love numbers could have been defined
following Schwarzschild's discovery of the black hole metric
\cite{Schw} it was only rather recently that works on the subject
appeared. \cite{FlanaganHinderer} computed the second Love number
for various models of neutron stars and discussed the possibility
of measuring it using gravitational wave detectors. Then in June
2009 three works appeared
\cite{DamourNagar,BinningtonPoisson,DamourLecian} which addressed
the black hole Love numbers, yet two issues remained open \bi

\item While both \cite{DamourNagar,BinningtonPoisson}
 independently found that a certain calculation yields zero
 contribution to the Love numbers, they disagreed on whether
 another contribution should be included, one arising from the effective field theory side and which was too
 difficult to compute at the time.
\footnote{\cite{DamourNagar} wrote ``The question of computing the
`correct' value of $k_l$ [the Love numbers] for a black hole is a
technically much harder issue which involves investigating in
detail the many divergent diagrams that enter the computation of
interacting point masses at the 5-loop (or 5PN) level.'' while
\cite{BinningtonPoisson} ``boldly proclaimed'' that the tidal Love
numbers of a black hole must be zero. Finally \cite{DamourLecian}
wrote ``there are subtleties inherent in any definition of the
multipole moments of BH's, so that there is currently no
unambiguous determination of the $k_l$ Love number of BH's.''}

\item \cite{DamourNagar} made an interesting speculation that
 a black hole may have no non-minimal world-line couplings whatsoever.
 \footnote{``This vanishing suggests, but in no way proves, that the effective action describing the gravitational
interactions of black holes may not need to be augmented by
nonminimal worldline couplings.'' \cite{DamourNagar}}

 \ei

We shall consider the determination of the Love numbers from the
point of view of the effective field theory approach to GR
\cite{GoldbergerRothstein1,Goldberger-Lect} including the ideas of
\cite{CLEFT-caged}.\footnote{See \cite{DamourFarese} for early
precursors of the EFT approach to GR.}
 We started working on this topic over two years ago but were held back by the above-mentioned
apparent ambiguity in their definition.

The strategy of this paper is to calculate the Love numbers for an
arbitrary space-time dimension following the idea of ``dimension
as a parameter of GR'', see \cite{kGL} and references therein
including \cite{kGL-asymp}. We shall discover an interesting
dependence which will also shed new light on the 4d issues.

We start in section \ref{sec:BHeff} by discussing the concept of
the black hole effective action and defining its static subsector.
We define the induced polarization constants. In section
\ref{sec:measure} we describe a concrete physical set-up for its
measurement together with the corresponding computation. In
section \ref{sec:microscopic} we perform the ``microscopic''
calculation, namely we find the appropriate
non-asymptotically-flat deformation of the \Schw solution and we
determine the Love numbers. The results are interesting and we
proceed to discuss them in subsection \ref{subsec:discuss}. In
section \ref{sec:div} we discuss divergences and their physical
meaning including counter-terms, classical RG flow and
cancellation of divergences in the EFT. In section \ref{sec:EFT}
we look at the calculation from an EFT perspective and reproduce
several results with Feynman diagrams. We end with a summary and discussion in section \ref{sec:summary}.

\section{The black hole effective action}
\label{sec:BHeff}

\begin{figure}[t]
\begin{center}
\includegraphics[width=10cm]{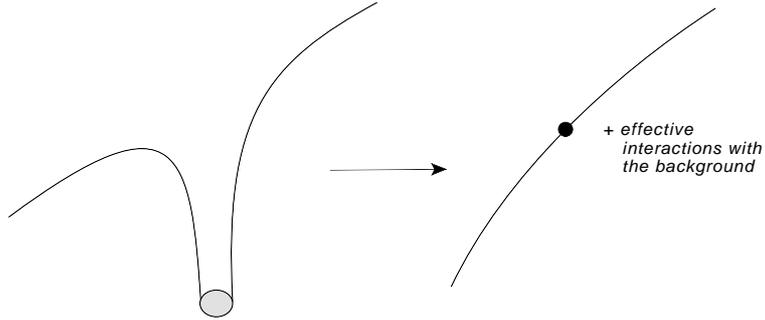}
\caption{The idea behind the black hole effective action: the full
black hole space-time geometry is replaced by a point particle
together with some effective interactions with its slowly varying
background.} \label{fig:idea}
\end{center}
\end{figure}

Consider a soliton of size $r_0$ moving in a background with
typical length scale $L \gg r_0$. Given the hierarchy of scales it
is natural to use the point particle approximation. However, this
approximation is certainly limited, as it does not retain almost
any of the object's physical properties, namely the finite-size
effects. To improve upon that one uses either a matched asymptotic
expansion between a near zone and a far zone, or an effective
field theory (EFT) approach which is the more modern tool (see for
example \cite{CLEFT-caged} which followed the EFT approach \cite{GoldbergerRothstein1}). The idea is illustrated in figure \ref{fig:idea}.

A black hole (BH) is a soliton of interest in Einstein's theory of
gravity. In \cite{GoldbergerRothstein1} a matching procedure was
outlined which produces the black hole effective action. Later
\cite{CLEFT-caged} defined it
essentially\footnote{\cite{CLEFT-caged} did not list the internal
degrees of freedom as they appear here -- those were later
discussed in \cite{GoldbergerRothstein2,dof}.} as follows \bea
 S_{full,eff}[x^\mu,e_A^\mu; Q_{int};g] &=&S_{EFT,eff}[x^\mu,e_A^\mu; Q_{int};g] \\
 :=S_{bulk}\(g_{BH}(x^\mu,e_A^\mu; Q_{int};g)\) \non
 &:=& S_{bulk}[g]+S_{BH}[x^\mu,e_A^\mu; Q_{int};g] \nonumber
 \eea
This definition means that the effective theory is built such that its effective action for a point particle in a given background,
is identical with the same effective action in the full theory with a full black hole metric inserted into the same background. Accordingly, given a space-time metric $g$, one must
solve for a black hole metric $g_{BH}$ which describes a black
hole (namely a horizon) embedded on an appropriate world-line given
by $x^\mu(\tau)$ in $g$ and having the prescribed spin and
internal degrees of freedom, $e_A^\mu, Q_{int}$ respectively. The full bulk action (Einstein-Hilbert + gauge fixing) is
evaluated on $g_{BH}$. The resulting expression has a (standard) bulk contribution and a contribution
localized on the world-line. The black
hole effective action is defined to be the latter.

We now wish to define the static sector of $S_{BH}$. In this
sector the background metric is static (namely time independent
and time reversal symmetric), the black hole location is fixed and
it cannot rotate nor excite its internal degrees of freedom.
Therefore in this sector \be S_{bulk}\(g_{BH}(g)\)=
S_{full,eff}[g] = S_{EFT,eff}[g] = S_{bulk}[g]+S_{BH,st}[g]
\label{def:S-BHst} \ee
 and the BH effective action depends only on the background (but not
 on any collective or internal coordinates). For self-consistency
the location of the black hole must be an equilibrium point of the
gravitational background, namely a static geodesic, which without
loss of generality we take to be the origin, 0.

In the static limit it is convenient to use the Non-Relativistic
Gravitational (NRG) field decomposition \cite{CLEFT-caged,NRG},
namely a temporal KK reduction, which replaces the Einstein field
$g_{\mu\nu}$ by three fields: the Newtonian potential $\phi$, the
gravito-magnetic vector potential $A_i$ and the spatial metric
$\gamma_{ij}$ according to \be
 ds^2 = e^{2 \phi}(dt - A_i\, dx^i)^2-e^{-2\phi/\hd}\,
 \gamma_{ij}\, dx^i dx^j \label{def:NRG} \ee
 where we conveniently denote \be
 \hd := d-3 ~. \ee
In the static limit $A_i=0$.

We define a Lagrangian by \be
 S_{BH,st}[\phi,\gamma] = \int dt\, \cl_{BH,st}[\phi,\gamma] ~. \ee
We expand $\cl_{BH,st}$ around flat space-time
$\gamma_{ij}=\delta_{ij}, \phi=0$. In general the leading term in
$S_{BH}$ is the universal point-particle action \bea
 S_{BH} &=& -m \int d\tau + \dots \non
 d\tau &:=& \sqrt{g_{\mu\nu}(x)\, dx^\mu\, dx^\nu} \eea
Therefore in the static sector \be
 \cl_{BH,st}^{(0)}  = -m  \ee

Perturbations around flat space-time are parameterized by $\phi, \sigma_{ij}$ where $\gamma_{ij}=\delta_{ij}+\sigma_{ij}$. Terms proportional to the equations of motion can be removed by a field re-definition. In particular  \be
 \del_{ii} \phi, ~ \del_i \phi|_0  \label{redund} \ee
 are redundant terms, the first being a linearized equation of motion in the bulk (assuming a vacuum space-time) and the second being the linearized geodesic equation (no gravitational field at 0).
We now wish to examine the leading terms beyond the point particle approximation, namely the finite-size corrections.
 There can be no terms linear in $\phi, \sigma$ due to the spherical symmetry (of the \Schw black hole)
 -- for instance a term linear in $\phi$ would represent a mass multipole for Schwarzschild.

The most general terms quadratic
in $\phi$ can be written as follows \bea
 \cl_{BH,st}  &=& -m + \frac{K}{2} \sum_{l=2}^{\infty} \lam_l \left| \del^{I_l} \phi \right|^2 +\dots \label{def:lam} \\
 \left|  \del^{I_l} \phi \right|^2 &:=& \frac{1}{l!} \sum_{I_l} \del^{I_l} \phi \del^{I_l} \phi  \eea
 where $I_l:=(i_1,\dots,i_l)$ is a multi-index and each $i$ runs
 over the spatial directions $i=1,\dots,d-1$ and $K$ will be defined shortly below in (\ref{def:K}).
We start with $\phi$ (rather than $\sigma_{ij}$) because the
leading finite-size effect for the post-Newtonian two-body
effective action comes from the quadrupole ($l=2$) term above.\footnote{In addition in 4d $\sigma_{ij}$ cannot fluctuate since in 3d
space Ricci flatness implies a flat space.}
 The sums starts with $l=2$: $l=0$ is absent since $\phi$ is only derivatively coupled, and $l=1$ is absent due to
 (\ref{redund}).
We refer to the constants $\lam_l$ as the induced gravito-static
polarization constants, or tidal Love numbers. Their normalization
here was chosen in a rather natural way, such that (for generic
$d,l$) \be
 Q_{I_l}=-\lam_l\, \phi_{I_l} \qquad \phi_{I_l}:=\del_{I_l} \phi|_0 \label{def2:lam} \ee
 where $Q_{I_l}$ is the mass multipole at the origin induced by the
background field multipole $\phi_{I_l}$.\footnote{Despite
considerable effort so far we were not able to find a clear enough
definition of the standard normalization of the tidal Love numbers
to allow comparison.} In particular, the factor K is chosen such
that the kinetic term for $\phi$ is $S \supset -K/2 \int \del_i
\phi\, \del_i \phi$. From(\ref{phi-action}) we have \be
 K := \frac{1+1/\hd}{8\pi G} \label{def:K} \ee
The sign convention is such that in weak (Newtonian) gravity
$\lam>0$. Most of this paper is devoted to computing $\lam_l$ for
all $d$. The dimensions of $\lam_l$ are \be
 \[ \lam_l \]= M\, L^{2l} = L^{\hd+2l} ~. \label{dim}\ee

Returning now to our title ``black hole stereotyping'' we see that
according to effective field theory it is useful to ``stereotype''
a BH when it is viewed from afar by a series of properties each
with a specific order.\footnote{Actually there is some freedom in
how we count the order. Apart from derivative counting we may also
assign dimensions to the fields $\phi,\sigma$. For instance in PN
$\phi$ should be considered order $m$ while $\sigma$ is order
$m^2$.} The mass is most relevant, followed by the spin which is
non-static and cannot be seen in (\ref{def:lam}). The next
property and the first finite-size one is $\lam_2$, and so on ...

\section{Measuring Love}
\label{sec:measure}

Here we translate the general definition of the Love numbers
$\lam$ (\ref{def:lam}) into an experimental measurement set-up and
a corresponding computation.

In the experimental set-up, following the description in the
introduction, one positions a non-rotating black hole at an
equilibrium point of a weak gravito-static field (such that the
typical distance for non-linear effects $L_{nl}$ is much larger
than $r_0$). The Newtonian potential $\phi$ is taken to be of the
form of a pure $l$-multipole, namely $\phi$ is a homogeneous
function of degree $l$ in $x^i$ the spatial coordinates and
satisfies the Laplace equation $\del_{ii} \phi=0$. One then
measures the induced mass $l$-multipole from the asymptotic form
of $\phi$, \footnote{Again at distances $r$ such that $r_0 \ll r
\ll L_{nl}$.}  and computes $\lam$ as the constant of
proportionality in (\ref{def2:lam}).

The computation starts with a background field $\phi(x)$
satisfying the flat space eq. of motion $\del_{ii} \phi =0$. As
usual we may perform a decomposition into spherical harmonics
$\phi = \sum_{lm} \phi_{lm}(r) Y_{lm}(\Omega)$ where in a general
dimension $m$ stands for a multi-index. A general solution to the
radial equation in flat space is of the form $\phi_l= A r^l + B
r^{-l-\hd}$. Imposing origin regularity we have \be
 \phi_l = \bphi\, r^l .\ee

Now we should obtain $\phi_{BH}$, the field in the presence of a
black hole which asymptotes to $\phi$ and is regular on the
horizon (for clarity we suppress now the index $l$). For $r \ll
L_{nl}$
we can write $\phi_{BH}$ as a linear combination \be
 \phi_{BH}(r) = \bphi \( \phi_1 (r) + \hlam \phi_2(r) \) \label{def:hlam} \ee
 where \bea
 \phi_1 &=& r^l \(1 + \dots \co \(\frac{m}{r^{\hd}} \) \) \non
 \phi_2 &=& r^{-l-\hd} \(1 + \dots \co \(\frac{m}{r^{\hd}} \) \) \label{phi-series} \eea
 and $\hlam$ is some constant.

Considering the equivalent definition of $\lam$ (\ref{def2:lam}),
and recognizing that the induced mass multipole at the origin
$Q_I$ is proportional to $\bphi \hlam$, the coefficient of
$\phi_2$,
 while $\phi_I$ is proportional to $\bphi$ we find that  \be
 \lam = N\, \hlam \label{def:N} \ee
 where the normalization constant is found to be (see appendix \ref{app:normalization}) \be
 N = 4 \pi \frac{\pi^{\hd/2}}{2^l\, \Gamma\(\frac{\hd}{2}+l\) } ~. \label{soln:N} \ee

We would like to make two comments about our procedure
(\ref{def:hlam}-\ref{soln:N}).

\noindent {\bf Possible ambiguity}. In the flat space theory
$\phi_1,\phi_2$ in (\ref{phi-series}) have no corrections.
However, in the presence of corrections the definition of $\phi_1$
 becomes ambiguous to adding a multiple of
$\phi_2$ whenever the powers in the two series (\ref{phi-series})
overlap, namely whenever $(2l+\hd)/\hd \in \IN \equiv
1,2,3,\dots$. We may express this as \be
 2 \hl +1 \in \IN \label{dimensional-condition} \ee
 where we defined \be
 \hl := \frac{l}{\hd} \equiv \frac{l}{d-3}  \label{def:lhat} \ee
Though in general $\hl$ need not be integral, in 4d $\hl$ is
nothing but $l$, and this ambiguity is essentially the one noted
in \cite{DamourLecian}.

A simple and effective solution is to observe that for generic
values of $l,d$ no such issue exists, and then the special values
can be treated as a limit. Alternatively, one must use the general
definition (\ref{def:S-BHst}) to separate the contributions to
$\hlam$ originating from two different EFT vertices
(\ref{def:lam}): the $m$ vertex and the $\lam$ vertex.

We note that in EFT language the condition
(\ref{dimensional-condition}) means nothing but that $\lam$ has a
dimension which is an integral power of $m$.

Later we shall find that the small parameter for the asymptotic
expansion (\ref{phi-series}) is actually $m^2/r^{2\hd}$, at least
in the background of a free scalar. Hence the condition for
possible ambiguity (\ref{dimensional-condition}) becomes more
restrictive, namely \be
 \hl + \half \in \IN \label{dimensional-condition2} ~. \ee

\presub {\bf Consistency with earlier terms of the EFT}. The
general definition of the static effective BH action
(\ref{def:S-BHst}) requires that the EFT action reproduces the
effective action of the full theory. In the present context it
means that corrections to $\phi_1$ (at least those which dominate
$\phi_2$) should be reproduced in the EFT by the $m$-vertex alone.
 Indeed the equation of motion involves the black hole metric which can be fully reproduced from the $m$-vertex
while the $\lam$-vertex reproduces the boundary condition (horizon
regularity). In section \ref{sec:EFT} we shall see several
examples for this consistency.

\section{Microscopic computation}
\label{sec:microscopic}

In our first computation we consider a minimally coupled scalar field $\psi$,
rather than the Newtonian potential $\phi$. We do that for several reasons
 (i) this case is simpler
 (ii) the gravitational case will turn out to be similar and
 (iii) this case actually describes the master field for tensor mode perturbations \cite{IshibashiKodama1}, 
 namely perturbations of the spatial metric in $d>4$.

 The action is \be
 S[\psi] = \half \int \left| \del \psi \right|^2 := \half \int \sqrt{-g} d^dx\, g^{\mu\nu}(x)\, \del_\mu \psi\, \del_\nu \psi \label{scalar-action} \ee
 We note that $K_\psi=1$ instead of (\ref{def:K}).
We wish to solve for $\psi$ in the background of a $d$ dimensional
black hole in \Schw coordinates \bea
 ds^2 &=& f\, dt^2-f^{-1}\, dr^2 - r^2 d\Omega^2_{d-2} \non
 f &:=& 1-\(\frac{r_0}{r}\)^\hd \label{Schw-metric} \eea
 where $r_0$ is the \Schw radius.

It is convenient to change coordinates $r \to X$ where \be
 X  := \frac{1}{r^\hd} \ee
namely $r_0^\hd\, X = 1-f $.\footnote{We record the standard
relation between the \Schw radius $r_0$ and the
 mass $m$, namely $r_0^\hd=16 \pi\, G m/\[(\hd+1)\Omega_{\hd+1}\]$.}

 Decomposing into (normalized) spherical harmonics the action becomes\footnote{Note that we consider the static limit,
 therefore the integral over time decouples and we suppress it in what follows.} \be
  S = - \frac{\hd}{2}\, r_0^{\hd} \int dX \[ f \left| \del_X \psi_l \right|^2 + \frac{\hl(\hl+1)}{X^2} \left|\psi_l \right|^2 \]
  \ee
 where $\hl$ was defined in (\ref{def:lhat}).
  The equation of motion is \be
  0= \[ \del_X f \del_X  - \frac{\hl(\hl+1)}{X^2} \] \psi_l ~. \label{eom1} \ee
We note that the equation of motion depends on $d,l$ only through
the combination $\hl$. Moreover, this static equation is analytic
in $X$ while the time dependent equation contains the term
$X^{-2/\hd} = r^2$ which is non-analytic for $d>5$. It would be
interesting to explain these simplifications.

\presub {\bf Solution.} The equation of motion (\ref{eom1}) has
three singularities all of which are regular: at the singularity,
the horizon and asymptotically. Therefore the solutions can be
expressed by the hypergeometric function. From hereon we use units
such that $r_0=1$.
The solution which is regular at the horizon is
\be \psi(X)=X^{-\hl}\, F(-\hl,-\hl,1;\; 1-X) =
 F(\hl+1,-\hl,1;\; -(1-X)/X)=P_\hl\(\frac{2-X}{X}\)
 \ee
 The second equality uses the Pfaff identity (\ref{Pfaff}) and $P_{\hl}$ are the Legendre polynomials.

In order to read $\hlam$ we must expand around $r=\infty$, namely
$X=0$. We do that using (\ref{expandX0}) and we use the Gamma
function identities (\ref{GammaId}) to simplify the expression,
finally arriving at \be
 \hlam_\psi =  \frac{1}{2^{4\hl+2}} \frac{\Gamma^2(\hl+1)}{\Gamma\(\hl+\half\) \Gamma\(\hl+\frac{3}{2}\)} \tan \pi \hl \; r_0^{2l+\hd} \label{pre-hlam-scalar} \ee
 where we restored units using powers of $r_0$.

\presub {\bf Isotropic coordinates}. At first sight dimensional
analysis (\ref{dimensional-condition}) allows for divergences when
$2\hl+1=1,2,\dots$, yet these occur only for $2\hl+1=2,4,\dots$.
This suggests that the small parameter can be taken to be $m^2$
rather than $m$.

Indeed, this is the case. The static Einstein-Hilbert action, when
expressed in terms of NRG fields (\ref{def:NRG}), reads
\cite{CLEFT-caged} \be
 S[\phi,\gamma_{ij}] = \frac{1}{16 \pi G} \int \sqrt{-\gamma} d^dx\, \[ -\(1+\frac{1}{\hd} \) \gamma^{ij}(x)\, \del_i \phi\,
 \del_j \phi + R[\gamma] \] ~ \label{phi-action} \ee
and enjoys the following parity symmetry \be
 \phi \to -\phi \qquad \gamma_{ij} \to \gamma_{ij} ~. \label{parity} \ee
 When supplemented by $m \to -m$ it is a symmetry also of the
relevant interaction term $-m \phi$. In gauges which respect this
symmetry, such as the isotropic and the harmonic gauges we may
conclude that $\gamma_{ij}=\gamma_{ij}(m^2)$ while
$\phi(-m)=-\phi(m)$. Moreover, $\psi$ couples only to
$\gamma_{ij}$ as can be seen by expressing its action
(\ref{scalar-action}) in the static case in terms of NRG fields
\be
  S[\psi] =  \half \int \sqrt{-\gamma} d^dx\, \gamma^{ij}(x)\, \del_i \psi\, \del_j \psi
  \label{scalar-action2} \ee
Therefore $m^2$ is indeed the small parameter for $\psi$,
thereby justifying the improved dimensional condition
(\ref{dimensional-condition2}).

The \Schw gauge which we used here (\ref{Schw-metric}) provides a
simple metric but does not respect this symmetry and therefore
this parity property was not manifest. It turns out that there is
a different representation of the black hole metric which
preserves the parity symmetry while providing equations of motion
which are quite similar to \Schw coordinates. These are the
isotropic coordinates \bea
 ds^2  &=& (f_-/f_+)^2 dt^2 - f_+^{4/\hd} (d\rho^2 + \rho^2 d\Omega^2) \non
 f_\pm &:=& 1 \pm (\rho_0/\rho)^\hd \label{def:isotropic} \eea
This metric is written in NRG form and satisfies the parity
property (\ref{parity}).

The coordinate transformation which takes us from isotropic to
\Schw is given by \be
 r^\hd = \rho^\hd \(1 + \( \frac{\rho_0}{\rho} \)^\hd \)^2 ~,
 \ee
In particular, the location of the horizon in isotropic
coordinates is at $\rho_0$ which is related to $r_0$ through \be
 \rho_0^\hd=\frac{1}{4} r_0^\hd ~.\label{def:rho0} \ee
The static action in isotropic coordinates now becomes
 \be
  S = -\frac{\hd}{2}\, \rho_0^{\hd} \int dZ f_+f_-\[ \left| \del_Z \psi_l \right|^2 + \frac{\hl(\hl+1)}{Z^2} \left|\psi_l \right|^2 \] ~,
  \ee
where \be Z=\({\rho_0\over \rho}\)^{\hd}~. \ee The solution to the
equation of motion which is regular at the horizon is given
by\footnote{This transformation is known in the mathematics
literature as a quadratic transformation of the hypergeometric
function, see for example \cite{BealsWong}.}
 \be
\psi(Z)=Z^{-\hl}\, F(1/2,-\hl,1;\; 1-Z^2) ~. \label{scalfullsol}
 \ee
In terms of $\rho_0$  our result (\ref{pre-hlam-scalar})
simplifies a bit and becomes \be
 \hlam_\psi =  \frac{\Gamma^2(\hl+1)}{\Gamma\(\hl+\half\) \Gamma\(\hl+\frac{3}{2}\)} \tan \pi \hl \; \rho_0^{\hd(2\hl+1)} \label{hlam-scalar} ~.\ee

Combining with (\ref{def:N},\ref{soln:N})
this completes the computation of the induced polarization
constants for the case of a scalar field background.

\subsection{Gravitational polarization}

Our starting point is the master equation for gravitational perturbations of scalar type in $d$ dimensions \cite{IshibashiKodama2} 
 which are the $d$ dimensional generalization of the Zerilli equation \cite{Zerilli} and describe perturbations associated with the Newtonian potential.
 In our notation the equation reads
\be
 \[\frac{d^2}{dX^2} + \frac{2(\hd-1)X+2}{\hd X(X-1)}\, \frac{d}{dX} + \frac{(\hd-1)X -(l-1)(l+\hd+1)}{\hd^2 X^2 (1-X)} \] Y(X) = 0
 \label{zerilli}
 \ee
 where we continue to suppress the index $l$, namely $Y \equiv Y_l$.
We change variables into \be
 \hy := r\, Y \ee
 and the equation reads \be
  \[ f \del_X^2 - 2 \del_X - \frac{\hl (\hl+1)}{X^2} \] \hy =0 ~. \label{eom:yh} \ee
This equation can be gotten from the following action \be
   S[\hy] = -  \int dX \[ f^2 \left| \del_X \hy \right|^2 + \frac{f\, \hl(\hl+1)}{X^2} \left|\hy \right|^2 \] \label{Yaction} \ee
As in the case of a minimally coupled scalar the equation is analytic in $X$ and depends on $l,d$ only through $\hl$.
We comment that this action encodes very economically the rather longer Zerilli potential,
 and that its form suggests that $\hy$ is the time component of a
contravariant vector, namely $\hy \equiv V^t$ for some vector $V$.

The solution is \be
 \hy(X)=X^{-\hl}\, F(1-\hl,-\hl,2;\; 1-X) = F(\hl+1,-\hl,2;\; -(1-X)/X) \label{soln:yh}
 \ee
 Expanding around $X=0$ we find \be \fbox{ $ \displaystyle
 ~~\hlam = - \frac{\Gamma(\hl)\Gamma(\hl+2)}{\Gamma\(\hl+\half\) \Gamma\(\hl+\frac{3}{2}\)} \tan \pi \hl\; \rho_0^{\hd(2\hl+1)}
  = -\(1+\frac{1}{\hl}\) \hlam_\psi $ } \label{main}
 \ee
 where $\rho_0$ was defined in (\ref{def:rho0}).

\begin{figure}[t]
\begin{center}
\includegraphics[width=10cm]{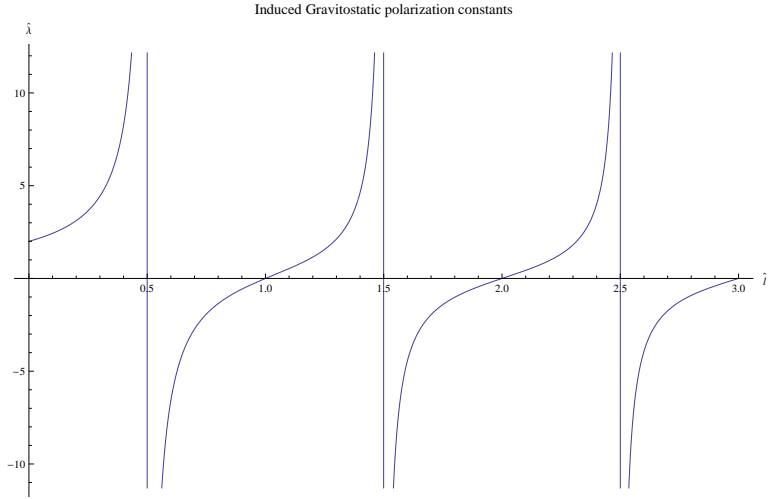}
\caption{A graph of the raw gravito-static Love numbers $\hlam$ of
a black hole as a function of $\hl\equiv l/(d-3)$ where $l$ is the
spherical harmonic index, and $d$ is the total space-time
dimension. Units are such that $\rho_0=1$ where $\rho_0$ is the
location of the horizon in isotropic coordinates
(\ref{def:rho0})}. \label{fig:hlam}
\end{center}
\end{figure}

The graph of $\hlam$ is given in figure \ref{fig:hlam}.

In order to relate the result for $\hlam$ (\ref{main}) with our
original definition (\ref{def:lam}) we must find the relation
between $\hy$ and the Newtonian potential $\phi$. When going to
the EFT we must take the flat space limit of $\hy$. Since there is
only one gauge invariant scalar the limit of $\hy$ and $\phi$ must
be proportional to each other, namely $\hy \to g(r)\, \phi$ for
some $g(r)$. Next we observe that in the flat space limit the
equation of motion for $\hy$ becomes $0=\triangle \hy$. Comparing
with the (flat space) equation of motion of $\phi$, namely
$0=\triangle \phi$ we conclude that $g(r)=const$ and actually we
may normalize our definition of $\hy$ such that $g(r)=1$.
Altogether we found that in flat space \be
 \hy \to \phi ~ \ee
(it would be nice to show this directly from the definition of
$\hy$, see also \cite{nRW,1dPert}).
 Therefore we can proceed and combine (\ref{main}) with the normalization factor (\ref{def:N},\ref{soln:N})
to obtain our main result  -- the induced gravito-static
polarization constants or a \Schw black hole.

\subsection{Discussion}
\label{subsec:discuss}

 We can now resolve the issues concerning 4d Love numbers which were listed in the
 introduction:
\bi
 \item 4d Love numbers (of electric type) vanish.

Indeed in 4d $\hd=1$, $\hl$ is always integral and $\tan \pi
\hl=0$. This settles a disagreement in the literature (described
in the introduction), validating the intuition of
\cite{BinningtonPoisson}. We note that in general
\cite{DamourNagar} were also right to consider an additional EFT
contribution, and indeed we shall find it to be non-zero for half
integral $\hl$. Yet in 4d it happens to vanish.

Our approach was able to avoid the danger of ambiguity in
definition since for generic dimensions none exists, and the 4d
limit is smooth. Thus it can be considered a dimensional
regularization. The 4d limit also serves as a first test of our
results.

The vanishing of Love numbers is quite surprising especially that
the black hole horizon is known to deform, see for example
\cite{dialogue,DamourNagar,DamourLecian,Vega:2011ue}. It implies a
certain infinite rigidity of the mass distribution of a 4d black
at least with respect to linearized gravito-static perturbations.

 \item Black holes do have non-minimal world-line couplings for
 $d>4$ (where $\hl$ could be non-integral). This settles negatively the speculation made in
 the abstract of \cite{DamourNagar}, at least for $d>4$.

\ei

The tangent factor in (\ref{main}) has several implications \bi
 \item $\lam$ vanishes whenever $\hl$ is integral.
 \item $\lam$ has a pole whenever $\hl$ is half-integral. This is
 related to a classical renormalization flow and will be discussed
 below. A second test of our result will come from a comparison of
 the residue of such poles against EFT divergences.
 \item $\lam(\hl)$ oscillates and changes sign. In particular it
 can be negative. The interpretation of a negative value is not
 clear at the moment. A negative compressibility of an elastic
 material implies an instability. However, a black hole cannot be
 divided and a negative specific heat, for example, does not imply
 an instability of a black hole, but rather of the black string
 \cite{GregoryLaflamme}. This instability is spherisymmetric
but we believe the during pinching this symmetry would be spontaneously broken. We speculate that negative
Love numbers are related to this last instability.

\item In 5d there are only zeros and poles.

\ei

We make some additional observations:  \bi

\item The factor in (\ref{main}) which includes the Gamma
functions has quite a smooth behavior for $\hl>0$, for large $\hl$ it approaches
1 and corrections in $1/\hl$ can be computed \be
 \frac{\Gamma^2(\hl+1)}{\Gamma\(\hl+\half\)
 \Gamma\(\hl+\frac{3}{2}\)} = 1 + \frac{3}{4\hl} - \frac{3}{32\hl^2} + \co \(\frac{1}{\hl^3}\) ~.\ee 
 This behavior suggests that a theory can be formulated in the $\hl \to \infty$ limit.

\item At integral values of $\hl$ the solutions are especially
 simple -- they are polynomials in $X$, since when either $a$ or $b$ are $0,-1,-2,\dots$ the hypergeometric series $F(a,b,c,;\; x)$ is finite.
A useful explicit formula is \be
 F(c+k,b,c;\; X) = \frac{\Gamma(c)}{\Gamma(c+k)} X^{1-c} \frac{d^k}{dX^k} X^{c+k-1}\, (1-X)^{-b} \ee
  for $k=0,1,2,\dots$
 \ei

\section{Diverging Love}
\label{sec:div}

In this section we examine more closely the case where $\hl$ is half integral and both Love
numbers (\ref{hlam-scalar},\ref{main}) diverge. The situation is
familiar in quantum field theory where it leads to RG flow,
and here it provides another example, quite pleasing, for classical RG
flow see \cite{GoldbergerRothstein1,dressed,GoldbergerRoss}.

\presub {\bf Cancellation of divergences}. While the gravitational
$\hlam$ diverges (\ref{main}), $\hy$, the horizon regular
wave-function, (\ref{soln:yh}) remains finite, of course. Looking
at the asymptotic expansion according to (\ref{expandX0}) and
substituting $a=1-\hl, b=-\hl, c=2$ for $\hy$ (\ref{soln:yh}) we
have \bea
 F(1-\hl,-\hl,2; 1-X) &=& \frac{\Gamma(2)\, \Gamma(2\hl+1)}{\Gamma(\hl+1) \Gamma(\hl+2)} F(1-\hl,-\hl,-2\hl; X) + \non
  &+& \frac{\Gamma(2)\, \Gamma(-2\hl-1)}{\Gamma(1-\hl) \Gamma(-\hl)} X^{2\hl+1}\, F(\hl+1,\hl+2,2\hl+2; X) \label{expand-hy}
 \eea
The divergence in $\hlam$ originates from  a divergence in the
factor $\Gamma(-2\hl-1)$ in the second summand. However, at the
same time the $c$ parameter in the first hypergeometric
function  becomes a negative integer $c_1=-2\hl \to -n$ which
implies a divergence in that function. Quite generally the
diverging part in such a case is given by
 \bea
 && F(a,b,-n+\eps;X) = \\ &=& X^{n+1}\frac{\Gamma(a+n+1) \Gamma(b+n+1)}{\Gamma(a) \Gamma(b) \Gamma(n+2)} \Gamma(-n+\eps)\, F(a+n+1,b+n+1,n+2;X) +
\co\(\eps^0\) \nonumber \eea
 where $n=1,2,\dots$ and $\Gamma(-n+\eps) = (-)^n/(n!\, \eps) + \co\(\eps^0\)$.
Using this relation with $a=1-\hl, b=-\hl, n=2\hl$ one confirms that the two divergences indeed cancel as $\eps \to 0$.

\presub {\bf Log terms}. As a byproduct of the cancellation of
divergences and the presence of dimensional constants a $\log(r)$
term is generated in $\hy(r)$. Indeed, as the divergences cancel
at order $\co\(\eps^{-1}\)$ the power law
$X^{2\hl+1}=X^{n+1-\eps}$ in the second summand (reaction term)
generates a finite $\co\(\eps^0\)$ $\log X$ term multiplied by the
residue of the pole. Hence the coefficient of the $\log$ term,
which is interpreted as the beta function, is directly related to
the coefficient of divergence (the pole residue in our dimensional
regularization).

The same phenomenon is seen also from a solution of the
differential equation of motion (\ref{eom:yh}) through the
substitution of an asymptotic power series $\hy=\sum_k \hy_k X^k$.
For half integral $\hl$ the resulting recursion relation at order
$k=2\hl+1$ becomes $0 \cdot \hy_{2\hl+1} = source \neq 0$, which
is known from the mathematical theory of differential equations to
signal the appearance of log terms.

\presub {\bf Finite, RG-flowing part}. Next we compute the finite
part. To do that, we must separate $\lam$ into a counter-term and
a finite part. The counter-term must cancel the EFT divergence.
The finite part depends of course on a choice of renormalization
scheme, and we use the standard MS (Minimal
Subtraction) scheme. As we saw $\log$ terms get generated and
accordingly the finite part depends also on the scale in which it
is evaluated. So altogether we must keep both the pole residue
and the finite part in $\lam$, but not higher terms in the $\eps$
expansion.

For convenience we turn to the case of a free scalar. Let us now
assume that $\hl=n/2$ is half integral, that is $n=1,3,5,...$.
Then the induced polarization (\ref{hlam-scalar}) diverges.
However, according to (\ref{scalfullsol}) the wave function $\psi$
is finite. As will be explained later, the divergent part of
induced polarization is cancelled by another divergence inherent
to the full solution (\ref{scalfullsol}). This means that
$\lambda_\psi$ is given by the sum of an infinite part and a
finite term which we now turn to compute. To separate these
contributions, we introduce the following definitions
 \be
 m=L^\epsilon m_r\, , \quad \epsilon=\hd-\hd_0\, ,
 \label{epsilon}
 \ee
where $L$ is an arbitrary length scale and $m_r$ has length
dimension -- $\hd_0$. Of course, $m$ must be $L$-independent,
therefore $m_r$ flows with $L$ according to
 \be
 L{dm\over dL}=0 \quad \Rightarrow \quad L{dm_r\over dL}=-\epsilon\, m_r ~.
 \ee
Substituting these definitions into (\ref{hlam-scalar}) and
expanding in $\epsilon$, yields\footnote{Note that $\rho_0$
depends on $\hd$, therefore one has to express it in terms of $m$
first and only then proceed with expansion in $\epsilon$.}
\begin{multline}
  \lam_\psi =   {N \, L^{\epsilon} \rho_0^{\hd_0(n+1)}\over 4\pi} {\Gamma^2\({n\over 2}\)\over
   \Gamma\({n+1\over 2}\)\Gamma\({n+3\over 2}\)}
  \\
  \times \Big( {2n\hd_0\over \epsilon}-2\,{n^2(n\hd_0-2)-n\over (\hd_0+1)(n+1)} +2 n^2 H_{n-1\over 2}
  -n^2\Big[ 2 H_{n\over 2}+\hd_0\log\pi-2\hd_0\log(L/\rho_0)\Big]
  \\
  +\hd_0\,n(n+1)\,\Psi(\hd_0/2+1)-{\hd_0\,n \over 2}\Psi(\hd_0(n+1)/ 2) +\mathcal{O}(\epsilon)\Big)~,
 \end{multline}
where $H_\al$ is the $\al$-th harmonic number and $\Psi(z)$ is the
digamma function. Notice that we should keep $L^\epsilon$ in front
of the above expression to maintain correct dimensions for
$\lambda_\psi$ in $(d+\epsilon)$-dimensional spacetime. In the
special case $l=2, \hd_0=4$ ($n=1$), we obtain
 \be
 \lambda_{\psi}\Big|_{l=2,\hd_0=4}={\pi^3\,L^{\epsilon}\rho_0^8\over 3}\[
 {1\over \epsilon}+{1 \over 2}\(\log{2\over\pi}+{7\over 60}-\gamma\)
 +\log{L\over \rho_0}+\mathcal{O}(\epsilon)\]~.
 \label{scalar7d-lam}
 \ee

The pole in $\epsilon$ corresponds to the divergence. In the EFT
approach this term represents the so-called counter term which has
to be introduced into the theory in order to make the observables,
e.g. the wave function $\psi$, finite. In contrast, the finite
piece in the expression for $\lambda_{\psi}$
represents the renormalized value of $\lambda_\psi$, i.e.
 \begin{multline}
  \lam_\psi^{rn}(L) =   {N  \rho_0^{\hd_0(n+1)}\over 4\pi} {\Gamma^2\({n\over 2}\)\over
   \Gamma\({n+1\over 2}\)\Gamma\({n+3\over 2}\)}
  \\
  \times \Big(2 n^2 H_{n-1\over 2} -2\,{n^2(n\hd_0-2)-n\over (\hd_0+1)(n+1)}
  -n^2\Big[ 2 H_{n\over 2}+\hd_0\log\pi-2\hd_0\log(L/\rho_0)\Big]
  \\
  +\hd_0\,n(n+1)\,\Psi(\hd_0/2+1)-{\hd_0\,n \over 2}\Psi(\hd_0(n+1)/ 2) \Big)~.
 \end{multline}
In particular, while $\lambda_\psi$ is certainly $L$-independent,
the induced polarization $ \lam_\psi^{rn}(L)$ exhibits a classical
RG flow
 \be
 \beta_{\lambda_\psi}=L{d \lam_\psi^{rn}(L)\over dL}= {2 N  \hd_0 \over \pi} {\Gamma^2\({n+2\over 2}\)\over
   \Gamma\({n+1\over 2}\)\Gamma\({n+3\over 2}\)}\rho_0^{\hd_0(n+1)}~.
 \ee

\section{EFT side}
\label{sec:EFT}

In this section we demonstrate and confirm the equivalence of the
EFT description  by mirroring some of the previous matching
calculations on the EFT side.

For convenience we consider the case of a free scalar field
(\ref{scalar-action}) and we choose the isotropic coordinates
(\ref{def:isotropic}).

Integrating out short scales of order $\rho_0$, leaves us with the
following effective action
 \bea
 S_{EFT, eff}(\sigma,\phi,\psi) =S_{bulk}(\sigma,\phi,\psi)+S_{BH,st}(\sigma,\phi)+ \half \sum_{l=2}^{\infty} \lam_l \left| \del^{I_l} \psi \right|^2 ~,
 \label{scalar-effact}
 \eea
where\footnote{For simplicity we set $G=1$ for the rest of this
section.} \be S_{bulk}(\sigma,\phi,\psi)=S[\psi]+{\hd(\hd+1) \over 64\pi}
  \int dx^{\hd+2}\sigma^{ \hd-4 \over 2}(\del \sigma)^2
 -{\hd+1\over 16\pi\hd} \int dx^{\hd+2}\sigma^{ \hd \over 2}
 (\del \phi)^2 ~,
\ee
and the indices are contracted with flat metric $\delta_{ij}$.
To simplify the gravity action (\ref{phi-action}), we set  $ \gamma_{ij}=\sigma\delta_{ij}$ for the background metric in isotropic coordinates and used the following
transformation rule for the $(d-1)$-dimensional curvature scalar
 \be
 \gamma_{ij}=e^{2\omega}\delta_{ij} \Rightarrow
 R[\gamma]=-e^{-2\omega}[2(\hd+1)\nabla^2\omega+\hd(\hd+1)\del_{i}\omega\del^{i}\omega]
 ~,
 \ee
where $\omega=\log(\sigma/2)$. Note that the full action is
quadratic in $\psi$ and symmetric to constant shifts,
$\psi\rightarrow \psi+\text{const}$. This explains why only
quadratic terms in the derivatives of $\psi$ are present in the
above effective action.

\begin{figure}[t]
\begin{center}
\includegraphics[width=11cm]{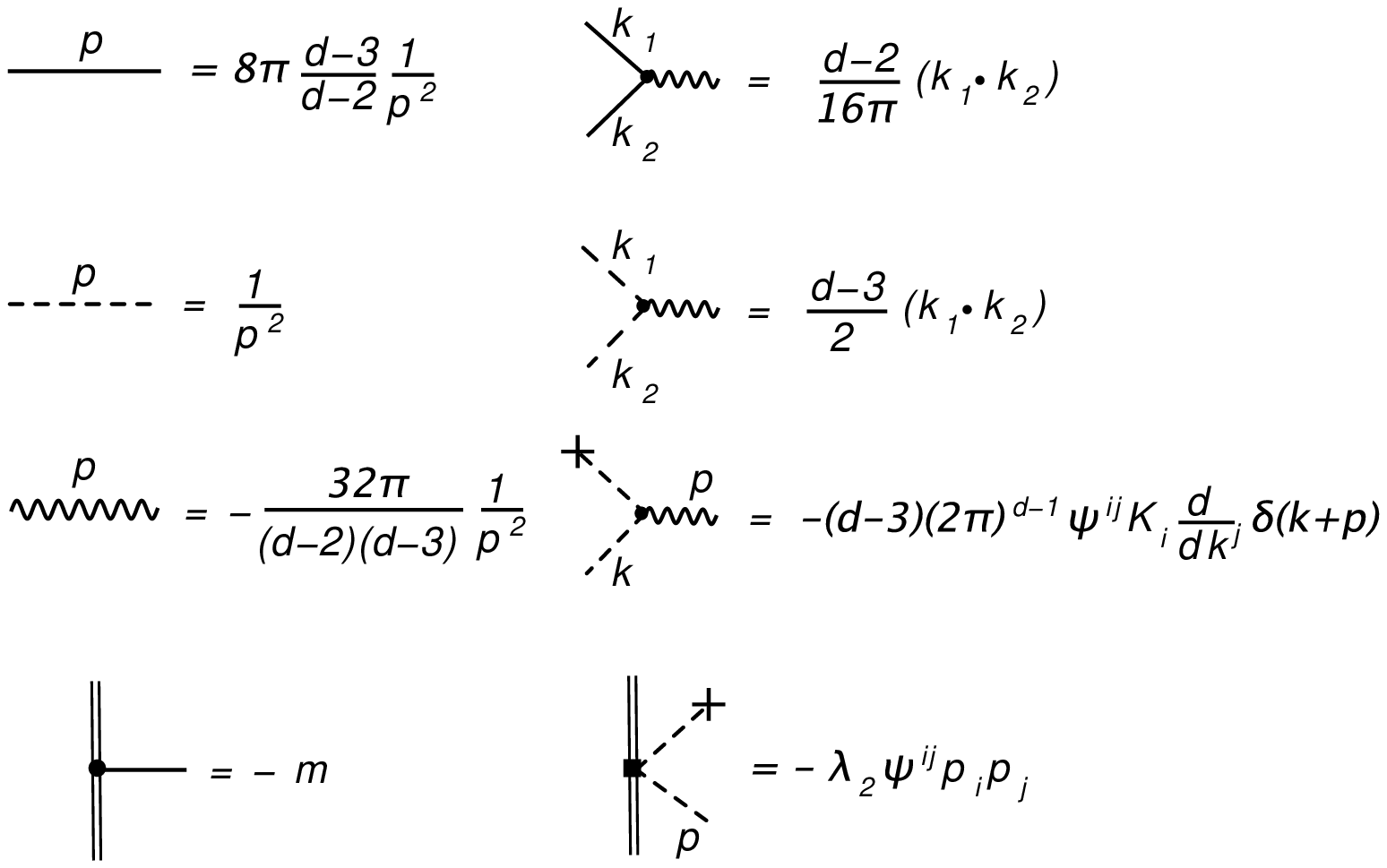}
\caption{All necessary diagrams to match $\lambda_2$ in isotropic
coordinates. Solid, dashed and wavy lines represent propagators of
$\phi,\psi$ and $\hat\sigma$ respectively. Cross indicates
insertion of $\bar\psi$. Wave numbers flow into the vertex.}
\label{fig:scrules}
\end{center}
\end{figure}

In what follows we match the quadrupole $\lambda_2$.
Introducing the following perturbation
 \be
 \psi\rightarrow \bar\psi+\psi\, \quad \bar\psi=\psi_{ij}x^i x^j , \quad Q^i_i=0
 \ee
into the effective action (\ref{scalar-effact}), we evaluate the
asymptotic expansion of the scalar field, $\psi_{EFT}$, and
subsequently match the result with the corresponding expansion of
the full solution (\ref{scalfullsol}). The Feynman rules can be
obtained by expanding (\ref{scalar-effact}) in the weak field
approximation, i.e., $\hat\sigma=\sigma-1,\phi,\psi\ll 1$. All
necessary rules for our needs are listed in figure
\ref{fig:scrules}. In what follows we use dimensional
regularization to evaluate various diagrams. All necessary
formulas to accomplish the computations can be found, e.g. in the
appendices of \cite{dressed}.

\begin{figure}[t]
\begin{center}
\includegraphics[width=11cm]{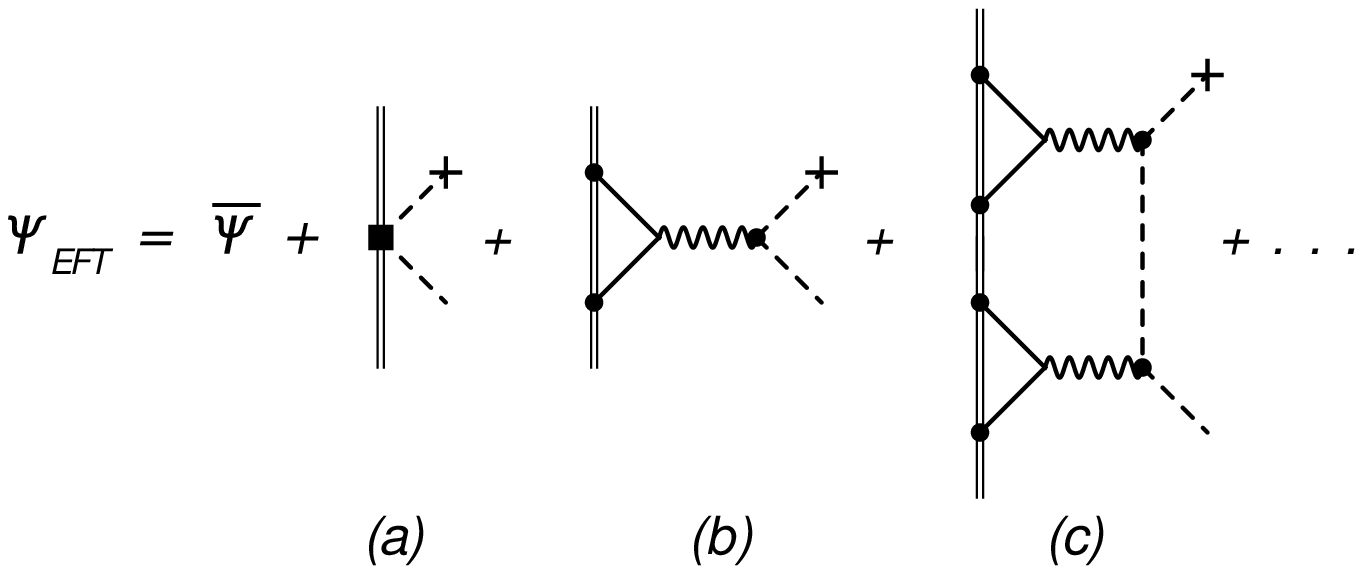}
\caption{Diagrams contributing to the asymptotic value of the
scalar field $\psi$.} \label{fig:scfield}
\end{center}
\end{figure}

The contribution of  the induced quadrupole $\lambda_2$ to
$\psi_{EFT}$ is shown in figure \ref{fig:scfield}(a) and is given
by
 \be
 \text{Fig.\ref{fig:scfield}(a)}={\lambda_2 \over \rho^{\hd+4} }  \, {\Gamma\big({\hd+4 \over 2}\big) \over \pi^{\hd/2+1}}\, \bar\psi~.
 \label{figA}
 \ee
For $\hd\geq5$ this contribution dominates all other possible
Feynman graphs, since according to (\ref{dim}) in this case
$\lambda_2\sim m^{\al}$ with $1<\al<2$, whereas as explained above
the small parameter of the theory is $m^2$. Expanding
(\ref{scalfullsol}) in $Z\ll 1$, yields
 \be
 \psi\Big|_{\hd\geq 5}=\bar\psi\( 1+\({\rho_0 \over \rho}\)^{\hd+4} {\Gamma^2(1+2/\hd) \over \Gamma(3/2+2/\hd)\Gamma(1/2+2/\hd)} \tan(2\pi/\hd)+\ldots\)~.
 \ee
Matching with the EFT result, we obtain
 \be
 \lambda_2=
 {\pi^{\hd/2+1} \Gamma^2(1+2/\hd) \over \Gamma(\hd/2+2)\Gamma(3/2+2/\hd)\Gamma(1/2+2/\hd)} \tan(2\pi/\hd) \rho_0^{\hd+4}~.
 \ee
This result agrees with (\ref{hlam-scalar}) upon substituting
$l=2$ and taking into account (\ref{def:N}).

\presub {\bf The divergence at $l=2,\, d=7$}. Let us now consider
$\hd=4$, namely $l=2$ in 7d.\footnote{The more general case of
$\hl=1/2$ with arbitrary $d$ is quite similar.} In this case we
should include the diagram in figure \ref{fig:scfield}(b), since
$\lambda_2\sim m^{2}$
 \be
 \text{Fig.\ref{fig:scfield}(b)}=\bar\psi\({\rho_0 \over \rho}\)^{2\hd}{\Gamma(2-\hd/2)\over \Gamma(3-\hd/2)}~.
  \label{figB}
 \ee
This diagram diverges in $\hd=4$.  Thus in the language of quantum
field theory, we have to renormalize the indefinite (bare)
parameter $\lambda_2$ such that $\psi$ becomes finite. We now
demonstrate this procedure.

Using (\ref{epsilon}) to expand figure \ref{fig:scfield}(b) in
$\epsilon$, yields
  \be
 \text{Fig.\ref{fig:scfield}(b)}=-{\bar\psi\over 4} \({\rho_0 \over \rho}\)^{8}\( {8\over\epsilon}+{44\over 5}-8\gamma-8\log\pi-16\log{\rho\over L} +\mathcal{O}(\epsilon)\)~.
 \label{figB2}
 \ee
where $\gamma$ is the Euler constant. To cancel the pole in
$\epsilon$, we redefine $\lambda_2$ as follows
 \be
 \lambda_2=L^{\epsilon}\rho_0^8\(\lambda_2^{(0)}+{\lambda_2^{(1)}\over \epsilon}+{\lambda_2^{(2)}\over \epsilon^2}+\text{other possible poles in $\epsilon$}\)~,
 \label{lamdec}
 \ee
where $\lambda_2^{(i)}$, $i>1$ are $\epsilon$-independent
constants adjusted to eliminate all poles in $\epsilon$ which
appear in $\psi$, while $\lambda_2^{(0)}$ depends on an arbitrary
scale $L$ and should be matched to agree with the full solution in
this case. This is the so called minimal subtraction (MS) scheme.

\begin{figure}[t]
\begin{center}
\includegraphics[width=7cm]{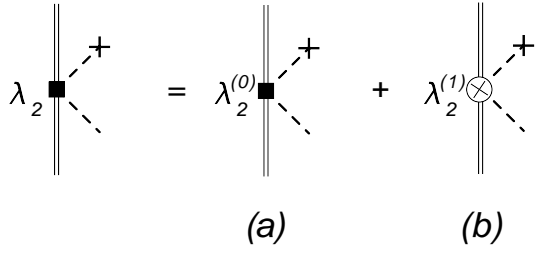}
\caption{Pictorial representation of decomposition of
(\ref{lamdec}). (a) represents the finite part, whereas (b)
corresponds to an infinite counter term.} \label{fig:counter}
\end{center}
\end{figure}

Pictorially, the above redefinition can be depicted as in figure
\ref{fig:counter}. Namely, one can think about it as decomposition
of the bare parameter $ \lambda_2$ into a finite piece
proportional to  $\lambda_2^{(0)}$ and an infinite part associated
with $\lambda_2^{(1)}$. In the QFT literature the latter is called
a counter term.

Substituting (\ref{lamdec}) into (\ref{figA}) and expanding in
$\epsilon$, yields
 \be
  \text{Fig.\ref{fig:scfield}(a)}={\bar\psi\over\pi^3} \({\rho_0 \over \rho}\)^{8}
  \Big[ {6\lambda_2^{(1)}\over\epsilon}+6\lambda_2^{(0)}+{11\over 2}\lambda_2^{(1)} -3\lambda_2^{(1)}\gamma-3\lambda_2^{(1)}\log\pi-6\lambda_2^{(1)}\log{\rho\over L} +\mathcal{O}(\epsilon) \Big]~.
 \ee
To cancel the pole in (\ref{figB2}), we choose
 \be
 \lambda_2^{(1)}={\pi^3\over 3}~.
 \ee
Hence,
 \begin{multline}
 \psi_{EFT}\Big|_{\hd=4}=\text{Fig.\ref{fig:scfield}(a)}+\text{Fig.\ref{fig:scfield}(b)}+\ldots
 \\=\bar\psi\[1+\({\rho_0 \over \rho}\)^{8}\({6\over \pi^3}\lambda_2^{(0)}-{11\over 30}+\gamma+\log\pi
 +2\log{\rho\over L}+\mathcal{O}(\epsilon) \)+\ldots\]~.
 \end{multline}

On the other hand, expanding the full solution (\ref{scalfullsol})
in the vicinity of $\hd=4$, we obtain
 \be
 \psi\Big|_{\hd=4}=\bar\psi\[1+\({\rho_0 \over \rho}\)^{8}\({4\log 2 -1 \over 4}
 +2\log{\rho\over \rho_0} \)+\ldots\]~.
 \ee
First thing to note is that EFT correctly reproduces the leading
$\log\rho$ term. Finally, matching the results, yields
 \bea
 \lambda_2^{(0)}(L)&=&\lambda_2^{(0)}\Big|_{L=\rho_0}+{\pi^3 \over 3}\log{L\over \rho_0}~,
 \non
 \lambda_2^{(0)}\Big|_{L=\rho_0}&=&{\pi^3 \over 6}\(\log{2\over\pi}+{7\over 60}-\gamma\)~.
 \label{lam7d}
 \eea
The corresponding classical RG flow is thus given by
 \be
 \beta_{\lambda_2}=L{d \lambda_2^{(0)} \over d L}={\pi^3 \over 3}~.
 \ee
All the above results are inherent in our master formula
(\ref{hlam-scalar}) as is evident from (\ref{scalar7d-lam}).

\presub {\bf Subleading corrections}. When $\hd=2,3$ there is no
need to compute new diagrams. Indeed, in this case $\lambda_2\sim
m^{\al}$ with $2<\al\leq 3$, therefore only diagrams in figures
\ref{fig:scfield}(a) and \ref{fig:scfield}(b) are relevant for our
needs. Using (\ref{figA}) and (\ref{figB}), we obtain
 \bea
 \psi_{EFT}\Big|_{\hd=2}&=&\bar\psi\[1+\({\rho_0 \over \rho}\)^{4}+{2\over \pi^2}{\lambda_2\over \rho^6}+\ldots\]~,
 \\
 \psi_{EFT}\Big|_{\hd=3}&=&\bar\psi\[1+2\({\rho_0 \over \rho}\)^{6}+{15\over 8\pi^2}{\lambda_2\over \rho^7}+\ldots\]~.
\eea Comparing with the asymptotic expansion of the full solution
(\ref{scalfullsol}), yields
 \bea
 \lambda_2\Big|_{\hd=2}&=&0
 \\
 \lambda_2\Big|_{\hd=3}&=&- {8\,\pi^2\sqrt{3}\,\Gamma^2(5/3)\over 15\,\Gamma(7/6)\,\Gamma(13/6)}\rho_0^7~.
 \eea
These results agree with (\ref{hlam-scalar}).

The last case to consider is $\hd=1$. In this case, the full
solution degenerates into a polynomial of fourth order in $Z$
 \be
 \psi\Big|_{\hd=1}=\bar\psi\[1+{2\over 3}\({\rho_0 \over \rho}\)^{2}+\({\rho_0 \over \rho}\)^{4}\]~.
 \label{sca4dfullsol}
 \ee

In particular, there is no need in any computation to fix
$\lambda_2$. Indeed, since as argued above, in isotropic
coordinates the small parameter  is $m^2$, whereas according to
(\ref{dim}) $\lambda_2\sim m^5$ in $\hd=1$, we conclude that there
are no diagrams with mass insertions which scale in the same way
as figure \ref{fig:scfield}(a). Furthermore, the full solution
does not contain odd powers of $m\sim\rho_0$. As a result, we
conclude that $\lambda_2$ must be zero to match the full solution
in this case. Of course, this result fits our general formula
(\ref{hlam-scalar}).

 It is instructive to show how the EFT reproduces the full solution (\ref{sca4dfullsol}). This requires computation of one additional diagram apart from what we have computed so far, i.e. figure \ref{fig:scfield}(c)
  \be
  \text{Fig.\ref{fig:scfield}(c)}=-{3(\hd-2) \over(\hd-4)(3\hd-4)} \, \bar\psi \,\({\rho_0 \over \rho}\)^{4\hd}~.
  \ee
For $\hd=1$ this diagram reproduces the last term in
(\ref{sca4dfullsol}). Note also that as expected it vanishes in
$\hd=2$. Indeed, in this case the full solution
(\ref{sca4dfullsol}) truncates at second order in $Z$, therefore
the overall contribution at order $Z^4$ must vanish.  As a last
test of this result we checked that it produces correct
contribution in $\hd=3$.

\section{Summary of results}
\label{sec:summary}

\noindent {\bf Results}. Our main result is the determination of
the induced gravito-static polarization constants, or
gravito-electric Love numbers, for any space-time dimension $d$,
given by (\ref{main}) together with the text below it and the
normalization factor (\ref{soln:N}).

We also computed the Love numbers for a free scalar field given
essentially by (\ref{hlam-scalar}). In $d>4$ these describe also
the tensor mode perturbations of the spatial metric.

\newpage
\presub {\bf Tests}. The results passed two tests \bi
 \item The 4d results are consistent with \cite{DamourNagar,BinningtonPoisson}. Their vanishing is dictated by the  $\tan \pi \hl$ factor.
 \item The residue of the pole at $\hl=1/2$ was compared with the EFT thereby confirming the overall normalization factor.
 \ei

\presub {\bf Main implications}. \bi
 \item We settled a disagreement in the literature whether the calculation described
 in \cite{DamourNagar} and \cite{BinningtonPoisson} implies a vanishing Love number or not.
 We find that in hindsight it indeed implies so and we supply the missing argument.
 This means that even though the horizons of 4d black holes do deform they are nevertheless ``infinitely rigid'' in this sense.
 \item We answered negatively (at least for $d>4$) a conjecture about the absence of non-minimal world-line couplings in the BH effective action.
 \item Our results for higher dimensions $d>4$ are unquestionably novel, as well as those for the free scalar field.
 \item We observed that the Love numbers are negative for certain ranges of parameters and interpreted that as indication for a novel non-spherical instability.
\ei
See more in the discussion subsection \ref{subsec:discuss}.

\vspace{1cm}

On the way we touched upon the following points of interest \bi
 \item We defined the static sector of the black hole effective action (\ref{def:S-BHst}).
 \item We defined $\hl$ (\ref{def:lhat}) and noticed that most of the analysis depends on $l,d$ only through it.
  \item We found quite a simple action  (\ref{Yaction})  which encodes the full Zerilli potential.
 \item We noted that there is an interesting limit to the theory as $\hl \to \infty$ where only the $\tan \pi \hl$ term survives in (\ref{main}).
 \item We provided a nice example of a classical RG flow, analyzed in sections \ref{sec:div},\ref{sec:EFT}.
\ei

\subsection*{Acknowledgments}

A significant part of BK's work was carried in the pleasant
environment of the workshop ``Gravity - New perspectives from
strings and higher dimensions''  (July 2011, Benasque). BK
appreciates the hospitality of the meeting ``the 6th regional
meeting on string theory'' (June 2011, Milos) and of G.
Sch\"{a}fer at Jena University (Aug 2011).

This research was partly supported by the Israel Science
Foundation grants no 607/05 and 812/11, by the German Israel
Cooperation Project grant DIP H.52, and by the Einstein Center at
the Hebrew University, though for most of the relevant time BK was
not supported by research grants. Research at Perimeter Institute
is supported by the Government of Canada through Industry Canada
and by the Province of Ontario through the Ministry of Research \&
Innovation.

\appendix

\section{Matching normalization constant}
\label{app:normalization}

\label{app:normalization}

In this appendix we evaluate the normalization constant $N$ (\ref{def:N}). This is done by solving the EFT equation of motion in the presence of the $\lam$ term (\ref{def:lam})
(obviously in flat space-time) and reading off the $\hlam$ coefficient defined in (\ref{def:hlam}).

The relevant part in the EFT action reads \be
 S \supset -\half \int \del_i \psi\, \del_i \psi + \half \lam_l \left| \del^{I_l} \psi \right|^2
 \ee
where $\left| \del^{I_l} \psi \right|^2$ is defined in (\ref{def:lam}). The following analysis does not change by a multiplication of the action by an overall constant.
The equation of motion is \be
 0= \triangle \psi + \frac{\lam_l}{l!} \sum_{I_l} \del_{I_l} \psi |_0\, \del_{I_l} \delta(-x) \label{eom:psiN} \ee
 Following (\ref{def:hlam}) we write the solution as \be
 \psi = \frac{1}{l!} \sum \psi_{I_l} x^{I_l} \( 1 + \frac{\hlam}{r^{2l+\hd}}\) \ee
 Substituting back into (\ref{eom:psiN}) we find that \be
  N= (-)^l \(c_0 c_1\)^{-1} \label{Nint} \ee
  where \bi
  \item $c_0$ is defined to be such that \be
  \triangle \(-\frac{c_0}{r^\hd}\) = \delta(x)
  \ee
  namely, $G_0=-c_0/r^\hd$ is a Green function satisfying $\triangle G_0 = \delta (x)$. We have \be
  c_0 = \frac{1}{\hd\, \Omega_{\hd+1}} = \frac{\Gamma\(\frac{\hd}{2}\)}{4 \pi^{\frac{\hd}{2}+1}} ~.
  \ee
  where $\Omega _{\hd+1}:=\mbox{Vol}\({\bf S}^{\hd+1} \)$. 
  \item $c_1$ is defined such that \be
  \del_I \frac{1}{r^\hd}=c_1 \frac{x^I}{r^{2l+\hd}}
  \ee
  where $I=(i_1,\dots,i_l)$ and all the indices $i_1,\dots,i_l$ are distinct. We find \be
  c_1 = (-2)^l \(\frac{\hd}{2}\)_l = (-2)^l \frac{\Gamma\(\frac{\hd}{2}+l\)}{\Gamma\(\frac{\hd}{2}\)}
  \ee
  \ei
   Substituting back into (\ref{Nint}) we obtain the requested quantity \be
   N = \frac{\hd\, \Omega_{\hd+1}}{2^l \(\frac{\hd}{2}\)_l} = \frac{\pi^{\frac{\hd}{2}+1}}{2^{l-2} \Gamma\(\frac{\hd}{2}+l\)}
   \ee

\section{Useful formulae}
\label{app:formulae}

We collect some useful facts for the microscopic calculation,
starting with the hypergeometric function, see for example \cite{BealsWong}.

The hypergeometric equation for $u=u(y)$ is \be
 y(1-y)\, u'' + [c-(a+b+1)y] u' - a b\, u =0 \ee
 The characteristic exponents are encoded in the Riemann P-symbol \be u(y) = P\(
 \begin{array}{ccc}
 0   & 1     & \infty \\
 0   & 0     & a \\
 1-c & ~c-a-b~ & b \end{array}; y \) \ee

The hypergeometric series is defined by \bea
 F(a,b,c;x) &:=& \sum_{k=0}^{\infty} \frac{(a)_k (b)_k}{(c)_k\, k!} x^k \non
 (a)_k &:=& \frac{\Gamma(a+k)}{\Gamma(a)} \equiv a \cdot (a+1)  \dots  (a+k-1) \eea

Pfaff's identity is \be
 F(a,b,c;X) = (1-X)^{-b} F\(c-a,b,c;\frac{X}{X-1}\) \label{Pfaff} \ee

Expansion around $X=0$ is given by \cite{BealsWong}(8.3.6) \bea
 F(a,b,c; 1-X) &=& \frac{\Gamma(c)\, \Gamma(c-a-b)}{\Gamma(c-a) \Gamma(c-b)} F(a,b,a+b+1-c; X) + \non
  &+& \frac{\Gamma(c)\, \Gamma(a+b-c)}{\Gamma(a) \Gamma(b)} X^{c-a-b}\, F(c-a,c-b,1+c-a-b; X) \label{expandX0} \eea

The Legendre polynomial is related to the hypergeometric function
through \be
 P_l(x) = F(-l,l+1,1;\frac{1-x}{2}) \label{Legendre-n-HGF} \ee

Finally we record some useful Gamma function identities \bea
 \Gamma(x) \Gamma(1-x) &=& \frac{\pi}{\sin \pi x} \non 
 \Gamma(x)\, \Gamma\(x + \half\) &=& 2^{1-2x} \sqrt{\pi} \Gamma(2 x) \label{GammaId} \eea

\end{document}